# Assessment of Microscopic Ion Beam Field Variation using Fluorescent Nuclear Track Detectors

*Alexander Neuholz*[1, 2, *], *Steffen Greilich*[1,2]

[1]*Division of Medical Physics in Radiation Oncology, German Cancer Research Center (DKFZ),*

*Im Neuenheimer Feld 280, D-69120 Heidelberg, Germany*

[2]*Heidelberg Institute of Radiation Oncology (HIRO), National Center for Radiation Research in Oncology (NCRO), Im Neuenheimer Feld, D-69120 Heidelberg, Germany*

**Abstract**
Fluorescent Nuclear Track Detectors (FNTDs) feature superior, submicron spatial resolution that allows for single particle track detection. However, when assessing particle fluence from the measured track positions, discrimination of actual fluence patterns from stochastic fluctuations due to spatial randomness in particle arrival can only be done at considerably lower resolution. This work quantifies the spatial limits of fluence-based dosimetry of (heavy) charged particles and presents the tools to detect deviations from homogenous fluence in measured data. It is found that deviations in fluence (and hence dose) on a percent level cannot be detected in a carbon beam on scales smaller than several tenths of a millimeter even when using dose levels of 1 Gy. For typical fluences measured with FNTDs, read-out area side-lengths should be larger than 0.2 mm to detect fluence differences of less than 5 %.

[*] Corresponding author.
*Tel.:* +49-6221-42-2633;
*E-mail address:* a.neuholz@dkfz.de;
*Postal address:* Department of Medical Physics in Radiation Oncology (E040), Research Group 'Ion Beam Therapy' (E040-8), German Cancer Research Center (DKFZ), Im Neuenheimer Feld 280, 69120 Heidelberg, Germany

# 1. Introduction

Radiotherapy with protons or heavier ions has gained increasing attention [Durante and Loeffler, 2010]. As for any kind of radiotherapy, accurate dosimetry is essential. In the case of ion beams, additional information on beam quality is often needed, for instance to calculate the relative biological effectiveness (RBE).

On this account, fluence-based techniques can complement dosimetry in addition to ionometric and calorimetric measurements [Karger et al., 2010]. Beside well-established methods such as plastic nuclear track detectors (e.g. CR-39 [Durrani, 2008, Sinenian et al., 2011]) or (silicon-based) semiconductor devices ([Eisaman et al., 2011, Renker and Lorenz, 2009]), Fluorescent Nuclear Track Detectors (FNTDs) together with confocal laser scanning [Akselrod and Sykora, 2011] are currently studied as fluence-based dosimeters [Klimpki et al., 2016].

The FNTD technique allows evaluating the properties of single particle tracks from three-dimensional microscope images containing information of the energy deposition as fluorescence signal intensity. While the spatial resolution of the microscope is in the submicron range, the effective resolution for the assessment of fluence $\Phi$, i.e.

$$\Phi = 1/A \cdot \sum_{i=1}^{N} |1/\cos\theta_i|, \qquad (1)$$

is limited by the finite scanned detector area $A$. $N$ represents the number of tracks detected across $A$, and $\theta$ their respective polar angles. Due to the spatial randomness in particle arrival, $N$ fluctuates in a Poisson fashion, i.e. with a standard uncertainty of

$$\sigma(N) = \sqrt{N}. \qquad (2)$$

On small scales, these fluctuations will eventually become the main uncertainty for fluence measurements, since $A$ is usually well known. The fluctuations can mask the patterns[†] the investigator is actually interested in, such as heterogeneities due to beam delivery, absorber characteristics or insufficient particle detection. The uncertainty in $N$ also directly impacts absorbed dose measurements, since

$$D = \Phi \cdot S/\rho \qquad (3)$$

with $S/\rho$ being the mass stopping power. Therefore, the inherent granularity of ion beams on small scales will affect any dosimetry method. Its impact is more pronounced the smaller the effective area of the dosimeter is, and the lower the dose and the higher the LET of the particles are. Thus, a limit can be derived from these parameters either for the maximum resolution / minimal spatial scale achievable for a given actual fluence or dose difference sought after. Alternatively, an estimate can be calculated for the smallest detectable (fluence) difference given a specific area of measurement. This area can be the size of an FNTD image, the cross-sectional area of an ionization chamber or the cell nucleus (see also [Rahmanian et al., 2017]).

---

[†] i.e. two-dimensional patterns of track spots over the lateral area of the FNTD perpendicular to the beam axis, which are referred to as "actual" in the following.

In this study, we quantify those limits for beam parameters commonly found in ion beam therapy and explore tools to extract actual patterns from fluence measurements with FNTDs as an example of use.

## 2. Materials and Methods

### *2.1. Detection limits*

If "Complete Spatial Randomness" (CSR) of particle arrivals, i.e. a homogeneous Poisson process, is assumed, the "Exact Poisson Test" [Hirji, 2006] can be used to calculate an estimate for the minimal spatial scale necessary to distinguish a given fluence difference $\Delta\Phi$. This two-sample test compares the rates (in our case fluences $\Phi$) of two Poisson distributed groups. The test returns the probability $p$ that their ratio is equal to one (meaning they belong to the same group) and a confidence interval for this ratio. $p$ can then further be compared to the significance level $\alpha$, that does not have to be defined in advance due to the "exactness" of the test.

We applied the Exact Poisson Test to situations representing typical FNTD measurements ($1 \cdot 10^6 - 5 \cdot 10^7$ cm$^{-2}$ for $^1$H, $^4$He and $^{12}$C beams), common clinical doses (1 mGy $-$ 1 Gy for $^1$H and $^{12}$C), and cellular area scales of interest (10 μm $-$ 100 μm).

### *2.2. Pattern analysis*

#### *2.2.1. Experiment*

A fluorescent nuclear track detector was irradiated at the Heidelberg Ion Therapy Center (HIT), placed in isocenter position on a PMMA plate free in air covering the whole detector area. We selected two different ion species for irradiation to enable additional discrimination of single tracks from first and second irradiation by the fluorescence amplitude in order to establish a ground truth on where the two parts are located exactly.

Initially, a single spot irradiation with $^4$He ions covering the whole FNTD was performed ($\Phi_{max} = 2.57 \cdot 10^6$ /cm$^2$, $E_{kin} = 220.51$ MeV/u, $LET_{Al2O3} = 5.56$ keV/μm). In addition, a part of the FNTD (1 mm of longer side) was irradiated with $^{12}$C ions ($\Phi = 2 \cdot 10^6$/cm$^2$, $E_{kin} = 88.83$ MeV/u, $LET_{Al2O3} = 92.02$ keV/μm, 5 x 5 cm$^2$ field size) to produce two groups of considerably different particle fluence. A 30 cm thick PMMA plate shielded the rest of the sample.

After irradiation, the radio-chromatically transformed color centers were read out with the FXR700RG automated reader (LANDAUER Inc., air objective 100x/ 0.95 NA, two current-mode APDs, 2D galvo-scanning, 640 nm diode laser [Akselrod et al., 2014]). The microscope provides a spatial resolution of 0.36 μm lateral and 1.7 μm axial over a total imaging area of 3 mm x 3 mm (i.e. 900 single image frames, each 6 slices in depth, scanning time 3 seconds per image[‡], 252 000 tracks in total). Further signal processing (track reconstruction and analysis) as described [Kouwenberg et al., 2016, Bartz et al., 2014] was done with the "FNTD package" [Greilich, 2016] for the R programming language [R Core Team, 2016]. Although the FNTDs store the information of nearly every particle (99.83% particle detection efficiency for carbon ions [Osinga et al., 2014]), the microscope readout and reconstruction of particle tracks can limit the number of detected particle tracks, especially for high fluences or large angles (e.g. two-digit percentage fluence uncertainty for Spread-out Bragg Peak (SOBP)).

---

[‡] High quality images used for LET and fluence determination are usually scanned with 10 seconds per image (each 504 x 504 px, 100 x 100 μm², 1 MB storage size) and with 21 slices in depth. This leads to an overall scan time of 210 seconds per frame and 21 MB disk storage. We reduced the scanning time and number of slices in depth for our FNTD sample, so that the fluence determination is still precise and at the same time the analysis reasonable fast (total scan time 15 h, 5.4 GB data).

*2.2.2. Tools*

Actual fluence patterns in the measured FNTD violate the pattern's CSR. To identify such patterns at scales above the minimal scale described in **Sec. 2.2.1**, a number of formal as well as informal tests from the R package "spatstat" [Baddeley et al., 2015] were investigated. In general, there are two explorable quantities:

Firstly, one can investigate the residuals of the pattern to assess first order effects, i.e. compare the measurement with Monte Carlo (MC) simulations of a pattern with similar properties (area size, number of particles and random distribution (CSR)). $\alpha$ can be adjusted by the number of simulations $n_{\text{sim}}$:

$$\alpha = 1/(n_{\text{sim}} + 1) \qquad (4)$$

We chose the "Pearson Residuals"[§] to express the discrepancy of the measured from the simulated patterns (four-panel plot, **Fig. 4** and **Fig. 5**). This enables to investigate spatial trends in x or y direction, caused by a fluence gradient, positioning outside the isocenter etc.

Secondly, correlation functions can describe the interaction between measured points (i.e. particle positions) based on the inter-particle distance $r$ (e.g. the commonly used Pair Correlation Function based or Besag's transformation of Ripley's *K*-function). For shorter distances, one can use the nearest neighbor or empty space function. All of these functions give insight into the second order violation of CSR, i.e. clustering, rejection, or regular pattern. MC sampling can generate an envelope for each function and compare these with the estimated function for a sample (e.g. Besag's *L*-function, **Fig. 6**).

For both, the residuals and the correlation functions, one can either investigate the graphical output with significance bands, to receive where (x- / y-direction or track spot distance $r$) the sample failed the CSR. Alternatively, one can use the $p$ values of these hypothesis tests, e.g. to compare a large number of samples. To calculate the $p$ values, the Maximum Absolute Deviation (MAD) test [Ripley, 1977], or for higher power the Diggle-Cressie-Loosmore-Ford (DCLF) test [Loosmore and Ford, 2006] are recommended [Baddeley et al., 2015].

---

[§] defined as raw residuals (difference estimated from simulated value) divided by the square root of the variance.

# 3. Results and discussion

## 3.1. Detection limits

### 3.1.1. Fluence-based calculations

**Fig. 1** summarizes the results of the Exact Poisson test and shows an inverse proportionality of minimal spatial scale $l_{\min}$ and square root of the dose

$$l_{\min} \sim 1/\sqrt{\Phi} \sim 1/\sqrt{D} \,. \tag{5}$$

This can also be derived from

$$\Delta\Phi/\Phi \sim 1/\sqrt{N} \tag{6}$$

(based on **Eq. (2)**) and **Eq. (3)**. The scaling factor for the inverse proportionality increases thereby for heavier ions due to their larger stopping power. This factor decreases for larger fluence differences and $\alpha$, illustrated in **Tab. 1 a)**. The table lists typical fluences for FNTD irradiations up to the maximal detectable fluence of $5 \cdot 10^7 \text{cm}^{-2}$ [Osinga et al., 2013]. Detection limits for these in the order of tens (or even hundreds) of micrometers.

### 3.1.2. Dose-based calculations

When expressing the fluence in terms of dose, lighter particles show less fluctuation in fluence at same dose due to their lower stopping power and hence larger $N$. **Tab. 1 b)** gives example figures for clinical situations. For instance, 1 % fluence deviations can only be measured on a two-digit micrometer scale at least. For Milligray measurements, this scale increases to a one-digit (two-digit) millimeter range for protons (carbons). More specifically, the measurement scale has to be 4.1 mm to gather an actual dose difference of 1 % for doses of 1 cGy. This is only slightly smaller than the sensitive diameter of a standard Farmer-type ionization chamber ($d = 6$ mm and considerably larger than the one of a pinpoint chamber ($d = 2$ mm).

### 3.1.3. Scale of a cell nucleus

In **Tab. 1 c)** the possible intervals for cell nucleus hits for clinical doses are listed. For small doses the interval of probable hits starts at zero, so that one has to increase the minimum dose to be sure to get at least one cell hit.

## 3.2. Pattern analysis

### 3.2.1. Particle distinction

**Fig. 2 a)** shows the fluence pattern of the FNTD, **Fig. 2 b)** the microscope signal intensity of track spots along the x-axis. In both illustrations, the fluence step around 340 µm is clearly visible. The second figure additionally allows distinguishing single tracks by particle type.

In **Fig. 3** the pattern is visualized with the minimal scale to resolve the fluence step (Exact Poisson Test: minimal scale of 24 µm for $\alpha = 5$ %) while avoiding fluctuations. A trend on the edges of the single images (regular peaks) is visible as well that may be caused by insufficient particle detection close to the image edges.

### 3.2.2. Toolbox outputs

The question remains if the toolbox can detect the fluence step and possible structures mentioned in **Sec. 3.2.1** as well.

On this account, the four-panel plot in **Fig. 4** was used to show that the part only irradiated with $^4$He agrees with CSR. This hypothesis holds except for the left part of the image, where more particles than expected occur. This is most likely to be caused by the $^{12}$C penumbra.

The four-panel plot in **Fig. 5** shows a fluence gradient along the x-axis on both bottom plots as one would expect due to the fluence step. In y-direction, the sample complies as expected with CSR due to the homogeneous irradiation in this direction.

# 4. Conclusion

FNTDs or any other detectors to study microscopic ion beam patterns are limited by stochastic fluctuations of particle arrival. At typical fluences of approximately $1 \cdot 10^7/\text{cm}^2$ measured with FNTDs, true differences of 10 % (1 %) can be resolved on a spatial scale of 0.1 mm (0.9 mm). This corresponds to 1 (9) microscope images of 100 x 100 µm² with 1000 particles detected in each. If additional parameters related to the tracks are assessed, e.g. "primary / fragment" or LET (in $n$ bins), the scales on which actual differences can be detected increase accordingly.

Although the maximum size of the FNTDs (4 x 8 mm²) allows for large readout areas, the scanning time per image has to be taken into account. For dosimetry, LET determination with FNTDs still suffers from uncertainties in the order of 20 % due to inter-detector fluctuation of sensitivity [Klimpki et al., 2016], which increases the dose uncertainty accordingly.

As fluctuations will affect any dosimetry in ion beams, the results suggest not expecting a resolution better than some tenths of a millimeter when looking for percent-effects even at dose levels of circa one Gy.

The tests applied to an FNTD in an exemplary manner were able to detect an actual, intentionally generated fluence step and furthermore gave insight to other effects (e.g. locally particle interactions of a few micrometer). This makes them a complementary tool to commonly used fluence / intensity histograms with the advantage, that only a two-digit number of particles is sufficient for the tests to work.

For dosimetry, in general the minimal scale is a good start to avoid measuring fluctuations. We recommend the use of the four-panel-plot with the established significance level of 5 % to get a first impression of the pattern's uniformity and apply more tests if irregularities occurred. Different CSR tests can lead to different results so that universally one alone is not sufficient.

## 5. Acknowledgements

We are grateful for the constructive input of the referees that greatly improved the manuscript. Furthermore, we like to thank Jeannette Jansen and Frank Orth for performing the experiments and fruitful discussions. We thankfully acknowledge the support of Stefan Brons with the experiments at HIT.

## 6. Funding

This research did not receive any specific grant from funding agencies in the public, commercial, or not-for-profit sectors.

## 8. Figures

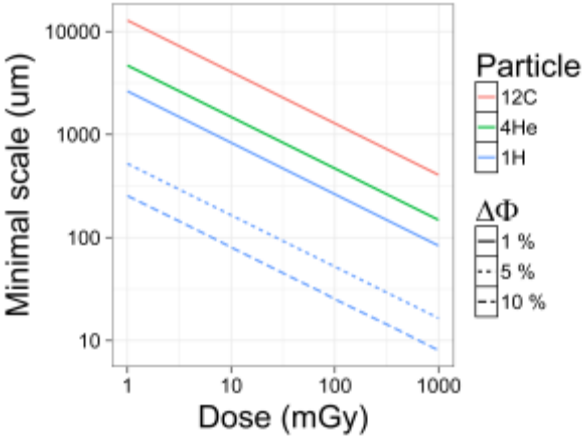

**Fig. 1** Minimal spatial scale ($\alpha = 5\ \%$) as a function of dose for three ion species ($^1$H 142 MeV/u, $^4$He 200 MeV/u, $^{12}$C 270 MeV/u) and three relative fluence differences $\Delta\Phi$. The spatial scale depends linear on the inverse of dose and fluence difference sought and increases with higher LET of the particle type.

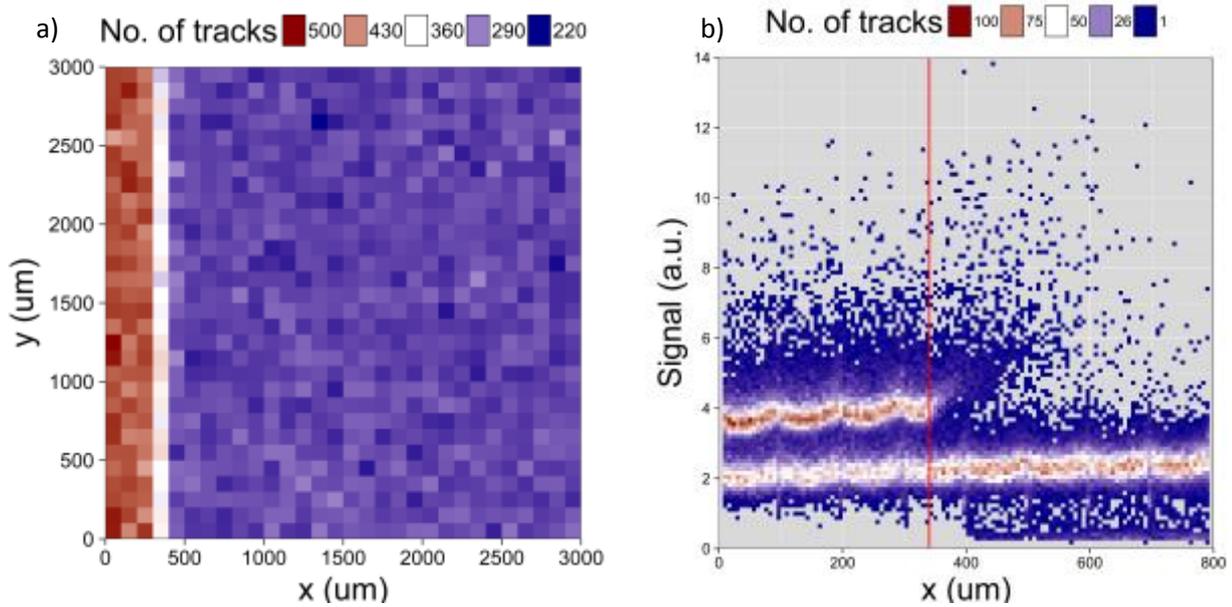

**Fig. 2** a) 2D histogram of spatial fluence. The FNTD was irradiated with Helium ($\Phi = 2.57 \cdot 10^6$ /cm$^2$) and partly (left side up to approximately 340 μm) with Carbon ($\Phi = 2 \cdot 10^6$ /cm$^2$). The step between the two fluences (in white) is clearly visible on the chosen scale (square size 100 x 100 μm$^2$; equal to the microscope image size).

b) Microscope signal intensity (maximum amplitude of the track spots) along the x-axis. $^4$He and $^{12}$C are clearly distinguishable due to their different LET. The intensity step around 340 μm (red line where partly irradiation with 12C ends) is clearly apparent and is used in addition to distinguish track wise between the particles' species

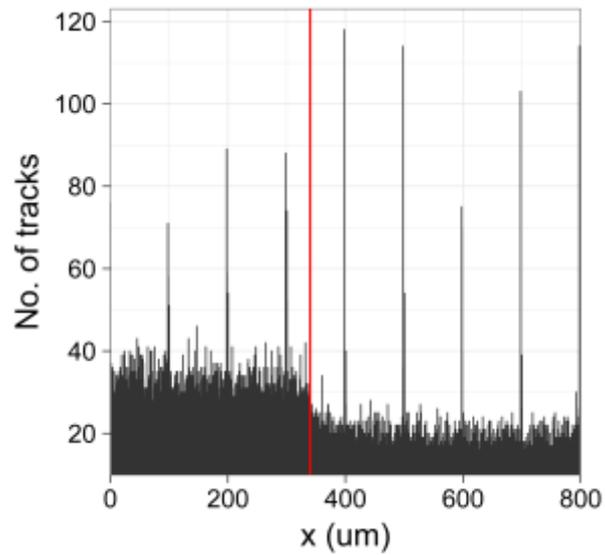

**Fig. 3** Number of tracks along the x-axis around the fluence step. The bin width was chosen to match the calculated scale (24 μm with Poisson Exact Test) to detect the fluence step (red line).

Not only the step but also a possible trend (fluence peaks at the end of every microscope image ($x = 100 \text{ μm}, 200 \text{ μm}$ etc.) is visible. The scale is sufficient to analyze the trend as well because the fluence difference between peaks and surrounding bins is higher than 50 %. The most likely reason for the peaks is an insufficient particle tracking close to the image borders. That should only be a local effect and average out on a slightly larger scale. Anyhow, it is good to know this exact scale, otherwise one would have used a larger scale to have better statistics but would have missed this trend.

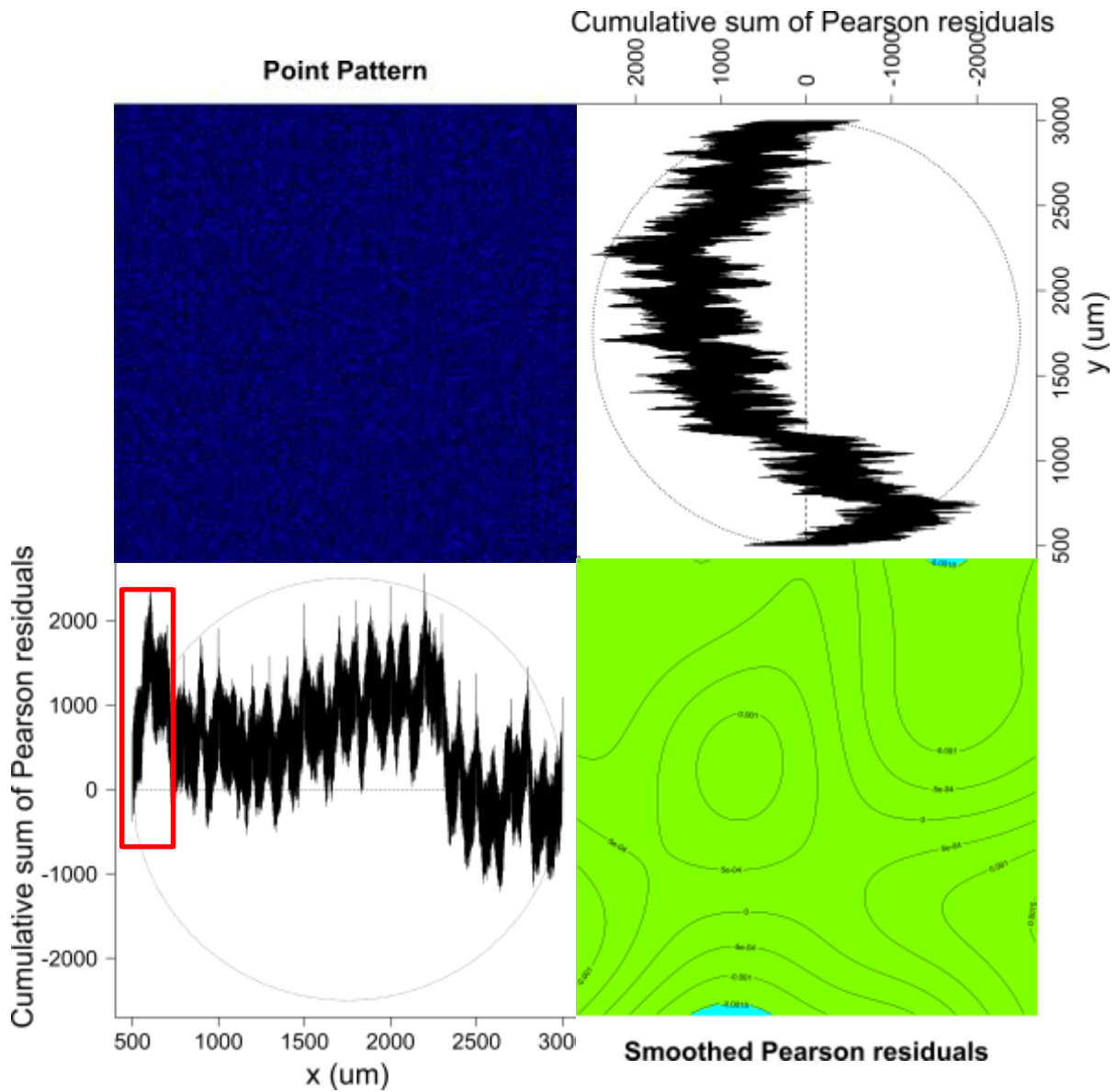

**Fig. 4** Four-panel plot of the FNTD's part only irradiated with $^4$He.

The upper left plot one shows the pattern itself, whereas the circles are at the track spots' positions and have the values of the positive residuals as diameter. The background color indicates the fitted point density of negative residuals; blue means underestimation and red overestimation.

In the top right and bottom left panel the cumulative sums of Pearson residuals along the two axes are shown (black lines). The dotted ellipses mark the model's pointwise two standard deviations.

The bottom right panel displays the Gaussian smoothed Pearson residual field.

This panel shows an 2.5 mm x 2.5 mm area only irradiated with $^4$He, where one would expect a homogeneous fluence. The smoothed field and the plot for the y-direction show no deviations from CSR. The bottom left plot has a small part where residuals are significant (red rectangle) at the very left, caused by the penumbra of the carbon field.

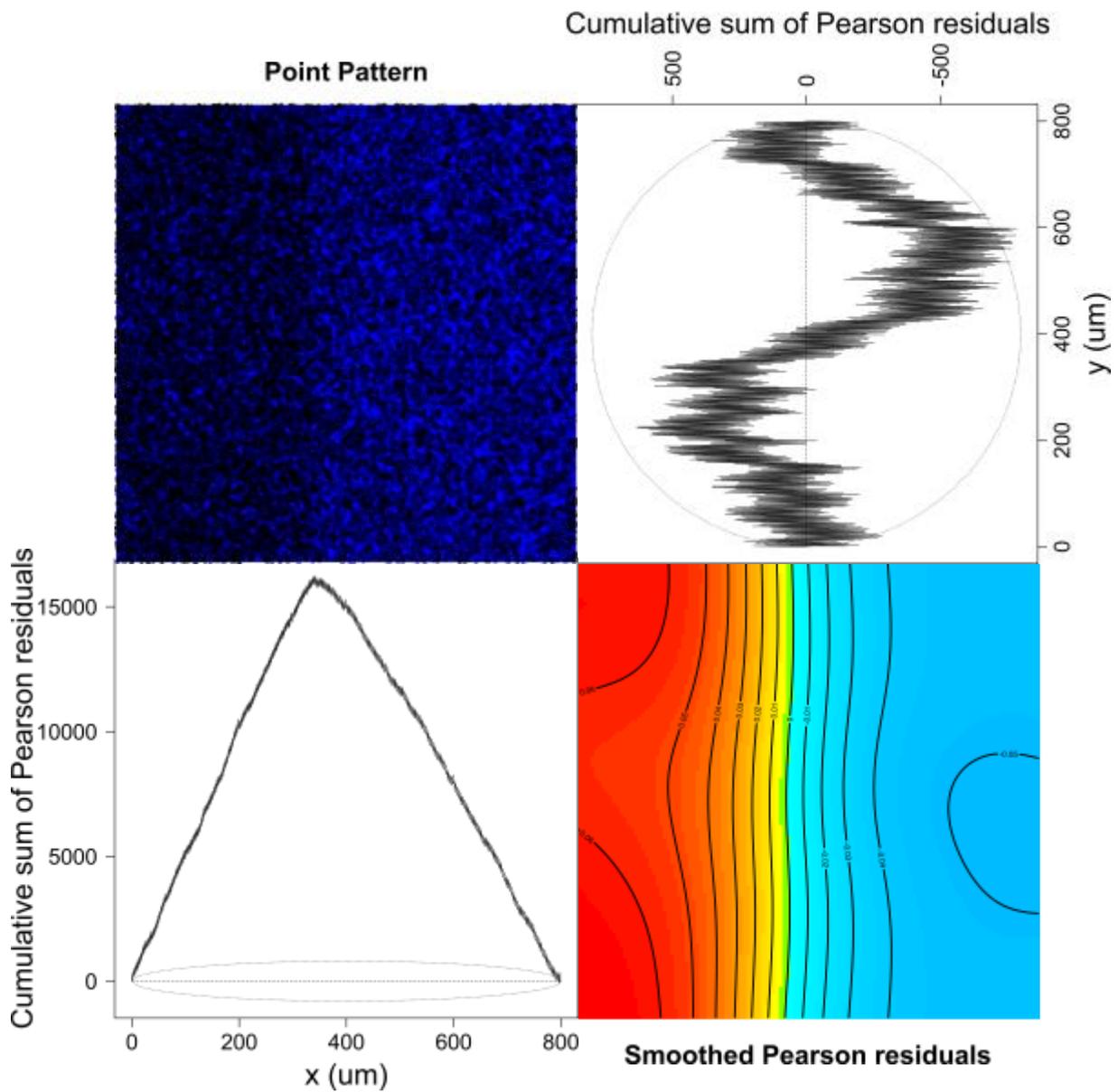

**Fig. 5** Four-panel plot of the fluence step.

In this panel the part of the sample (800 μm x 800 μm) with the fluence step (around $x = 340$μm) is displayed. The bottom left plot shows the expected fluence gradient in x direction clearly outside the significance bands. The density plot shows this gradient as well. In y-direction, there is no visible deviation from CSR.

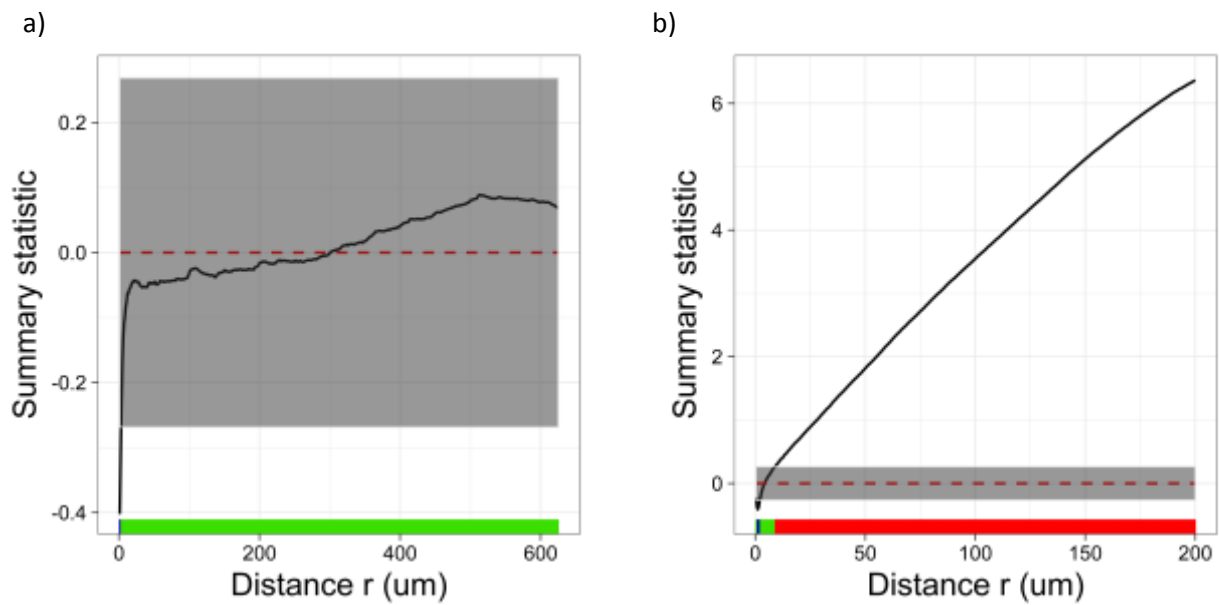

**Fig. 6** Besag's $L$-function for two parts of the analyzed FNTD. The maximum for the distance $r$ between track spots is defined as a fourth of the scanning area's side length. The envelope, displayed in grey with the red dotted line as mean value, is compared with the function for the sample (black line).

a) shows the part only irradiated with $^4$He (size 2.5 mm x 2.5 mm). The sample complies with CSR over a wide range of distances. Only distances of a few micrometers are significant (generally one would speak of repulsion/ underestimation) which could be caused by the definition of a critical distance between tracks in the particle tracker routine. This leads to a filtering out of tracks below this distance.

b) is a smaller area around the fluence step (size 0.8 mm x 0.8 mm). It shows except for very small distances (= local area around track spots) an overestimation of particles compared to the simulation caused by the step for distances over 10 μm. Below this distance, meaning for small inter-particle distances, the fluence step is not relevant and the sample complies with CSR. Again the underestimation caused by the particle tracker is visible.

## 9. Tables

**Tab. 1** a) Minimal spatial scale for typical FNTD fluences. Exemplarily fluence differences sought to detect are listed for two significance levels.

b) Scales for $^1$H and 12C at clinical doses are listed ($\alpha = 5\ \%$). A factor of 100 decrease in dose thereby increases the scale by a factor of 10.

c) Cell nucleus hits for prescribed doses ($\alpha = 5\ \%$). The first number indicates the needed hits for the demanded dose. The interval displays the possible number of hits. Especially for small doses, the relative deviation from the demanded hits can be quite problematic.

| a) | | | | | |
|---|---|---|---|---|---|
| Significance level $\alpha$ | Fluence difference $\Delta\Phi$ | Particle fluence $\Phi$ (cm$^{-2}$) | | | |
| | | $1 \cdot 10^6$ | $5 \cdot 10^6$ | $1 \cdot 10^7$ | $5 \cdot 10^7$ |
| 1 % | 1 % | 3.00 mm | 1.60 mm | 1.10 mm | 0.51 mm |
| | 10 % | 0.30 mm | 0.16 mm | 0.11 mm | 0.05 mm |
| 5 % | 1 % | 2.80 mm | 1.20 mm | 0.87 mm | 0.39 mm |
| | 10 % | 0.27 mm | 0.12 mm | 0.08 mm | 0.04 mm |

| b) | | | | | | |
|---|---|---|---|---|---|---|
| Fluence difference $\Delta\Phi$ | Proton dose | | | Carbon dose | | |
| | 1 mGy | 1 cGy | 1 Gy | 1 mGy | 1 cGy | 1 Gy |
| 1 % | 2.60 mm | 0.83 mm | 0.08 mm | 13.00 mm | 4.10 mm | 0.41 mm |
| 10 % | 0.25 mm | 0.08 mm | 0.01 mm | 1.20 mm | 0.39 mm | 0.04 mm |

| c) | | | | | | |
|---|---|---|---|---|---|---|
| Nucleus diameter | Proton dose | | | Carbon dose | | |
| | 1 mGy | 1 cGy | 1 Gy | 1 mGy | 1 cGy | 1 Gy |
| 10 μm | 1** [0, 78] | 9 [3, 25] | 862 [784, 949] | 1 [0, 78] | 1 [0, 78] | 36 [22, 59] |
| 100 μm | 86 [63, 118] | 862 [784, 949] | 86240 [85430, 87060] | 4 [1, 20] | 36 [22, 59] | 3640 [3480, 3810] |

---

** Values below one particle were rounded to one to fulfill the required discrete character of the Poisson distribution.